\documentclass[%
aip,
amsmath,amssymb,
reprint,%
]{revtex4-1}

\usepackage{graphicx}
\usepackage{dcolumn}
\usepackage{bm}

\usepackage[utf8]{inputenc}
\usepackage[T1]{fontenc}
\usepackage{mathptmx}
\usepackage{etoolbox}

\makeatletter
\def\@email#1#2{%
	\endgroup
	\patchcmd{\titleblock@produce}
	{\frontmatter@RRAPformat}
	{\frontmatter@RRAPformat{\produce@RRAP{*#1\href{mailto:#2}{#2}}}\frontmatter@RRAPformat}
	{}{}
}%
\makeatother
\begin{document}
	
	\preprint{AIP/123-QED}
	
	\title{Comment on "Rotating Spin and Giant Splitting: Unoccupied Surface Electronic Structure of Tl/Si(111)"}
	\author{A. F. Campos, K. Wang, A. Tejeda}
	\affiliation{Laboratoire de Physique des Solides, CNRS, Universit{\'e} Paris-Saclay, 91405 Orsay, France.}
	
	\date{\today}
	
	\begin{abstract}

	\end{abstract}
	
	\maketitle
	
	Rashba effect in 2D systems is extensively studied nowadays due to spintronics applications. The Letter \cite{stolwijk2013rotating} studies the fundamentals of spin-orbit interaction in 2D systems. Experimental evidence is claimed for the rotation of the spin polarization vector in Tl/Si from an in-plane Rashba polarization at $\overline{\Gamma}$ to the surface normal at $\overline{K} (\overline{K'})$ valleys. These results are possible thanks to the single setup that could measure spin-resolved inverse photoemission (IPES) with in- and out-of-plane sensitivity. This Comment clarifies that (i) when considering the full data set in the Letter, the in-plane polarization does not vanish at the valleys, (ii) the Letter does not explain that the out-of-plane data are not real measurements, in the sense that they are derived by considering the fulfillment of a theoretical symmetry or from an unspecified data treatment.

\begin{figure}[ht!]
	\includegraphics[width=0.5\textwidth]{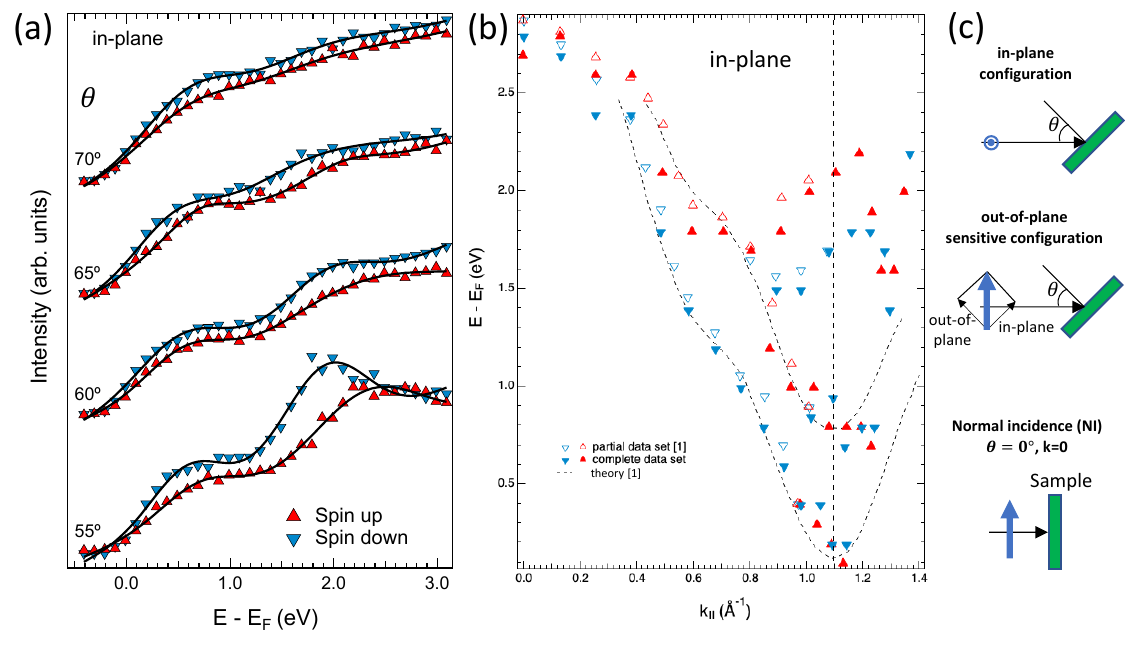}
	\caption{(a) Spectra near $\overline{K}$ for in-plane polarization from Ref. \onlinecite{stolwijk2013rotating}. Spin up/down spectra have different intensities, exhibiting features (peaks and shoulders) at different energies for spin up/down. (b) E vs $k_{||}$ along $\overline{\Gamma}\overline{K}$ with sensitivity to the in-plane polarization. The dispersion is derived from Fig. 2a in Ref. \onlinecite{stolwijk2013rotating}. Open triangles correspond to the partial dataset represented in Fig. 3a in Ref. \onlinecite{stolwijk2013rotating}. Filled triangles are derived from the full dataset in the Letter (from all the Fig. 2a spectra). Filled triangles are obtained from the second derivative of the spectra and follow the same trend as the experimental points displayed in the Letter, obtained from spectra fitting.  The full data set shows a non-negligible in-plane component around $\overline{K}$.  (c) Experimental geometries for the measurements. (Top) In-plane configuration. (Middle) In the configuration with some out-of-plane sensitivity, the polarization of the electron beam has non-vanishing in- and out-of-plane projections. The pristine out-of-plane component is not measured straightforwardly. (Bottom) Both at in- or out-of-plane configurations, at $\theta=0^{\circ}$, the beam reaches the sample at normal incidence (NI) with the spin perpendicular to the incidence, i.e. with in-plane sensitivity. The Letter presents out-of-plane results for this configuration (Fig. 2c ($\theta=0^{\circ}$) and Fig. 3b (k=0) in Ref. \onlinecite{stolwijk2013rotating}).}
	\label{fig:C1}
\end{figure}
	
	The Letter claims in-plane polarization vanishing around $\overline{K}$ and "completely out-of-plane spin-polarized valleys in the vicinity of the Fermi level". These claims are supported by the absence of points near the $\overline{K}$ valley in Fig. 3a of Ref. \onlinecite{stolwijk2013rotating}. This is at odds with the spectra around $\overline{K}$ ($55^{\circ} \le \theta \le 70^{\circ}$) showing differences in spin up/down (Fig. 2a of Ref.\onlinecite{stolwijk2013rotating}, from which Fig. 3a is extracted), i.e. a non-vanishing in-plane polarization. The difference in intensity and in energy for spectral features in spin up/down spectra is more visible in Fig. \ref{fig:C1}a, which just magnifies the spectra in the Letter. More importantly, the points associated with the spectral features showing the non-vanishing in-plane polarization are absent in Fig. 3a in Ref. \onlinecite{stolwijk2013rotating}, suggesting that the in-plane component vanishes at $\overline{K}$. Note that both panels in Fig. 2 in Ref. \onlinecite{stolwijk2013rotating} have the same angular range (0 to 70$^{\circ}$) and when $\theta$ is converted to $k$, the same wavevector range should be found in Fig. 3 panels. However $k^{max}_{\bot}<1.2$ \AA$^{-1}$ (Fig. 3b) while $k^{max}_{||}<1.1$ \AA$^{-1}$ (Fig. 3a), i.e. before $\overline{K}$. Fig. \ref{fig:C1}b here displays the in-plane dispersion when the full data set is considered (to be compared to Fig. 3a in the Letter), making explicit the in-plane component at $\overline{K}$. Some of these points have negligible spin splitting, other no. None of these is presented in the Letter, that displays experimental points with negligible spin splitting, away from $\overline{K}$.

In addition to the partial representation of the data set, it is required an explanation about the extraction of pristine out-of-plane results from experiments (Fig. \ref{fig:C1}c). After the Letter, the authors stated that their setup allowed to obtain out-of-plane results through the input of symmetry expectations or extra information from other experiments\cite{stolwijk2014rotatable}, without details of the procedure applied in the Letter for Tl/Si. In none of these situations, out-of-plane results will be direct measurements.

If the out-of-plane results have been deduced from a symmetry, which is the evoked symmetry? Expected symmetries have often not been fulfilled\cite{Park2018, Wu2013}. What is the experimental evidence that the expected symmetry applies in Tl/Si? In any case, even if it was experimentally demonstrated that the symmetry is fulfilled, a total vanishing of the spectral intensity can be absent\cite{Mulazzi2009}. The out-of-plane component must be measured. If the out-of-plane is measured thanks to the input from other experiments, the data treatment has to be specified by the authors. We just highlight here that comparing spectral intensities for measurements considering the macroscopical rotation of the electron source is delicate. There are many factors affecting intensities, as for example (a) the macroscopic rotation of the source displaces the electron beam from the optical axis and it must be realigned\cite{stolwijk2014rotatable}, (b) the beam polarization changes its direction with respect to surface orbitals and (c) sample and photocathode freshness evolve with time in the long IPES measurements, affecting also spectral intensities. The Letter concludes though on constant spin-integrated intensity for measurements at the same angle.
		
In summary, conclusions on the spin polarization vector rotation and the completely out-of-plane spin-polarized valleys should be reviewed by considering the full data set measured by the authors. Moreover the results do not correspond to the direct measurement of the out-of-plane polarization. Reliable results require sources allowing to measure in- and out-of-plane spectra without making assumptions \cite{Campos22a, Campos22b, Campos22c}. 
\bibliography{aipsamp}
		
\end{document}